\documentclass[a4paper]{jpconf}
\usepackage{graphicx}
\usepackage{citesort}

\begin{document}

%%%%%%%%%%%%%%%%%%%%%%%%%%%%%%%%%%%%%%%%%%%%%%%%%%%%%%%%%%%%%%%%%%%%%
%%%%%%%%%%%%%%%%%%    Head of the paper                %%%%%%%%%%%%%%
%%%%%%%%%%%%%%%%%%%%%%%%%%%%%%%%%%%%%%%%%%%%%%%%%%%%%%%%%%%%%%%%%%%%%

\title{Ab-initio calculation of the vibrational modes of 
  SiH$_4$, H$_2$SiO, Si$_{10}$H$_{16}$, and Si$_{10}$H$_{14}$O}

\author{Katalin Ga\'al-Nagy, Giulia Canevari, and Giovanni Onida}

\address{
  European Theoretical Spectroscopy Facility (ETSF),
  CNR-INFM  and Dipartimento di Fisica,
  Universit\`a degli Studi di Milano,
  via Celoria 16,
  I-20133 Milano, Italy 
}

\ead{katalin.gaal-nagy@physik.uni-regensburg.de}

\begin{abstract}
We have studied the normal modes of hydrogenated and oxidized
silicon nanocrystals, namely SiH$_4$ (silan), H$_2$SiO (silanon),
Si$_{10}$H$_{16}$ and Si$_{10}$H$_{14}$O. The small clusters
(SiH$_{4}$ and H$_2$SiO) have been used for convergence tests and
their bondlengths and frequencies have been compared with experimental
and theoretical reference data. For the large clusters
(Si$_{10}$H$_{16}$ and Si$_{10}$H$_{14}$O) we have investigated the
vibrational density of states where we have identified the
oxygen-related spectral features. The vibrational modes have been also
analyzed with respect to the displacement patterns. The calculations
have been carried out within the density-functional and
density-functional perturbation theory using the local-density
approximation.
\end{abstract}

%%%%%%%%%%%%%%%%%%%%%%%%%%%%%%%%%%%%%%%%%%%%%%%%%%%%%%%%%%%%%%%%%%%%%
%%%%%%%%%%%%%%%%%%    Body of the paper                %%%%%%%%%%%%%%
%%%%%%%%%%%%%%%%%%%%%%%%%%%%%%%%%%%%%%%%%%%%%%%%%%%%%%%%%%%%%%%%%%%%%

%
%%%%%%%%%%%%%%%%%%%%%%%%%%%%%%%%%%%%%%%%%%%%%%%%%%%%%%%%%%%%%%%%%%%%%
%
\section{Introduction}
Silicon nanostructures have important applications in microelectronics
due to the downscaling of optoelectronic devices. Because of this, the
optical, electronic, and vibrational properties of these systems are
of large interest. For example, photoluminescence of quantum wires has
been discovered \cite{Can1990}. The investigation of silicon
nanocrystals is another step in the direction of the development of
nanostructured devices.  Although the electronic and optical
properties of oxidized and non-oxidized nanocrystals have been studied
extensively \cite{Gat2005,Lup2003b,Lup2003a,Oss2006,Syc2005}, much
less is known about their vibrational properties. Nevertheless, the
production and operation of devices is carried out at room temperature
where the vibrations of the crystals play an important role. They can
enhance adsorption processes as well as they can influence the optical
properties of the systems.

Si$_{10}$H$_{16}$ and Si$_{10}$H$_{14}$O are good models for the study
of (oxidized) miniaturized semiconductors, since they are simple and
they can be studied with fully ab-initio total energy
calculations. Thus, we have studied the vibrational properties for
these systems as prototypes for non-oxidized and oxidized
nanocrystals. Besides the influence of the dynamics of the
nanocrystals to chemical processes, their vibrational frequencies can
also be utilized for the characterization of the oxidized clusters due
to the signature of the oxygen in the vibrational density of
states. In a first step we have studied smaller systems (silan and
silanon) to assess the numerical convergence. Furthermore, for silan
and silanon experimental and theoretical reference data exist which
can be used for comparison.

This article is organized as follows: after a short description of the
method employed in our calculations, we present the results for silan
and silanon (Sect.~\ref{SectSmall}), where the convergence tests, an
analysis of the displacement patterns, and a comparison with
experimental and theoretical results are shown. Then, we investigate
Si$_{10}$H$_{16}$ and Si$_{10}$H$_{14}$O (Sect.~\ref{SectBig}). Here, we
discuss again some convergence issues before going to the final
results for the vibrational density of states, which have been analyzed
with respect to oxygen-related features and their displacement
patterns. Finally, we summarize and draw a conclusion.

%
%%%%%%%%%%%%%%%%%%%%%%%%%%%%%%%%%%%%%%%%%%%%%%%%%%%%%%%%%%%%%%%%%%%%%
%
\section{Method}\label{SectMethod}
All calculations have been carried out with the ABINIT package
\cite{abinit}. It is based on a plane-wave pseudopotential approach to
the density-functional theory (DFT) \cite{Hoh64,Koh65}. We have
employed norm-conserving pseudopotentials in the Troullier-Martins
style \cite{Tro91}. The exchange-correlation energy is described
within the local-density approximation (LDA) \cite{Per81,Cep80}. The
phonon frequencies have been calculated utilizing the density-functional
perturbation (DFPT) scheme \cite{Gon97,Gon97b}. Since we apply a
plane-wave method to an isolated systems we have used the supercell
method and the $\Gamma$ point only in the {\bf k} point sampling. The
atomic positions have been relaxed till the residual forces have been
less than $0.5~{\rm mHa/a_B}$.

%
%%%%%%%%%%%%%%%%%%%%%%%%%%%%%%%%%%%%%%%%%%%%%%%%%%%%%%%%%%%%%%%%%%%%%
%
\section{Silan and Silanon}\label{SectSmall}
\begin{figure}[t]
\begin{minipage}{14pc}
\includegraphics[width=5.0pc]{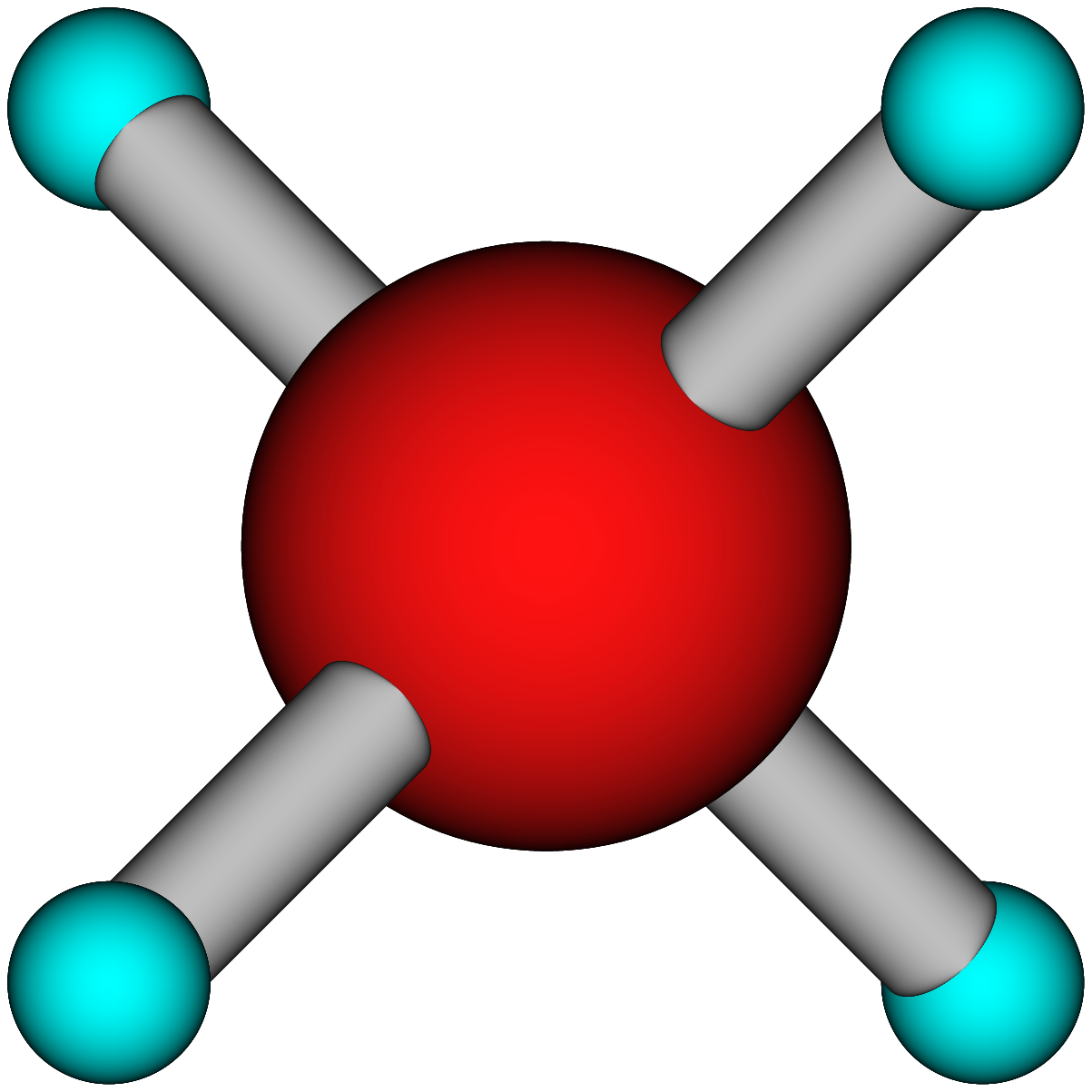} \hspace{3pc}
\includegraphics[width=5.0pc]{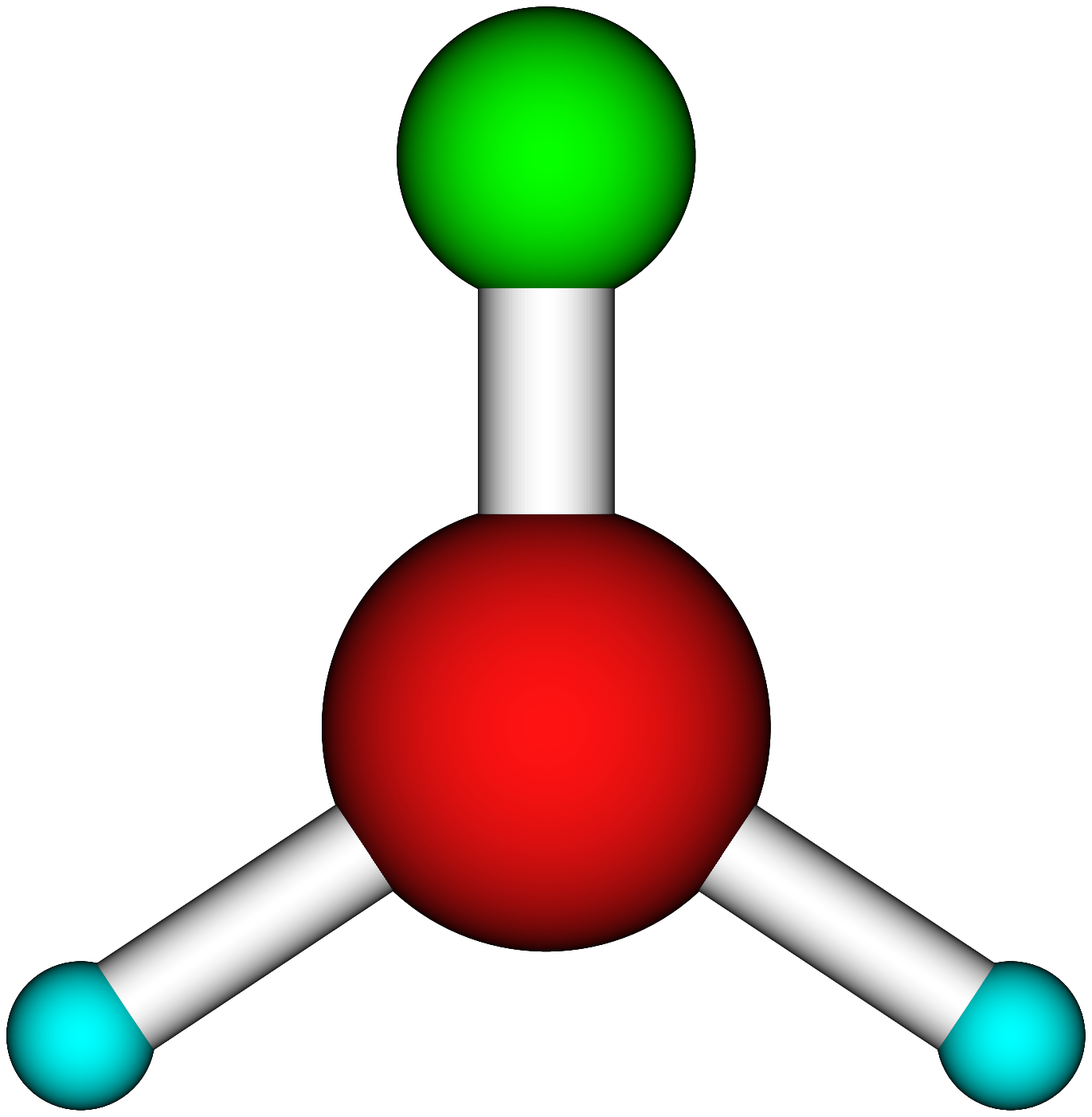}
\end{minipage}\hspace{2pc}%
\begin{minipage}{20pc}
\caption{\label{StructSmall} Silan (left) and silanon (right). Silicon
  atoms are drawn with large dark-red spheres, hydrogen atoms with
  small light-blue ones, and oxygen atoms with medium-sized green
  ones.}
\end{minipage} 
\end{figure}

In a first step we have studied silan (SiH$_4$) and silanon (H$_2$SiO)
as the smallest oxidized and non-oxidized silicon clusters. The atomic
structure is shown in Fig.~\ref{StructSmall}. Both systems are highly
symmetric, with space groups $T_d$ and $C_{2v}$, respectively. This
yields in parts to a degeneracy of the vibrational frequencies.

The small molecules have been utilized for the convergence tests
(Sect.~\ref{SectSmallConv}), since the investigation of the big
crystals is more time consuming. However, also the displacement
patterns for the various vibrational modes have been analyzed
(Sect.~\ref{SectSmallEig}), and the resulting frequencies have been
compared with reference values (Sect.~\ref{SectSmallExp}).

%
%%%%%%%%
%
\subsection{Convergence tests}\label{SectSmallConv}
Applying a plane-wave approach together with a supercell method, the
two main convergence parameters in the calculation are the
kinetic-energy cutoff ($E_{\rm cut}$), which determines the number of
plane waves used in the expansion, and the size of the (cubic) supercell,
meaning the amount of vacuum around the isolated cluster. The latter
parameter should be chosen as small as possible in order to reduce the
computational effort, but large enough in order to avoid an
interaction of the neighboring clusters, since periodic boundary
conditions are employed.

\begin{figure}[t]
\begin{minipage}{12.5pc}
\includegraphics[width=12.5pc]{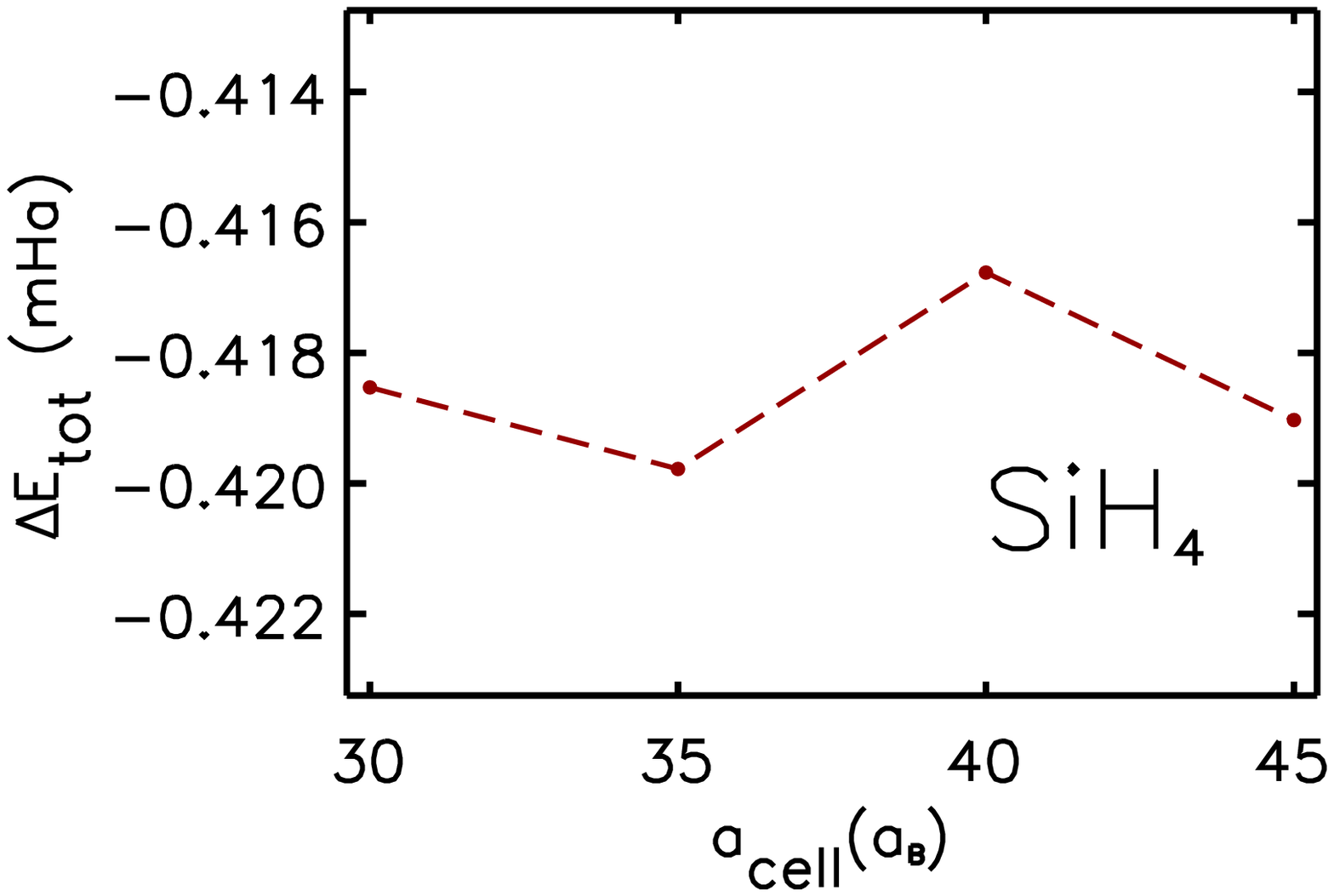}
\end{minipage}%\hspace{1pc}%
\begin{minipage}{12.5pc}
\includegraphics[width=12.5pc]{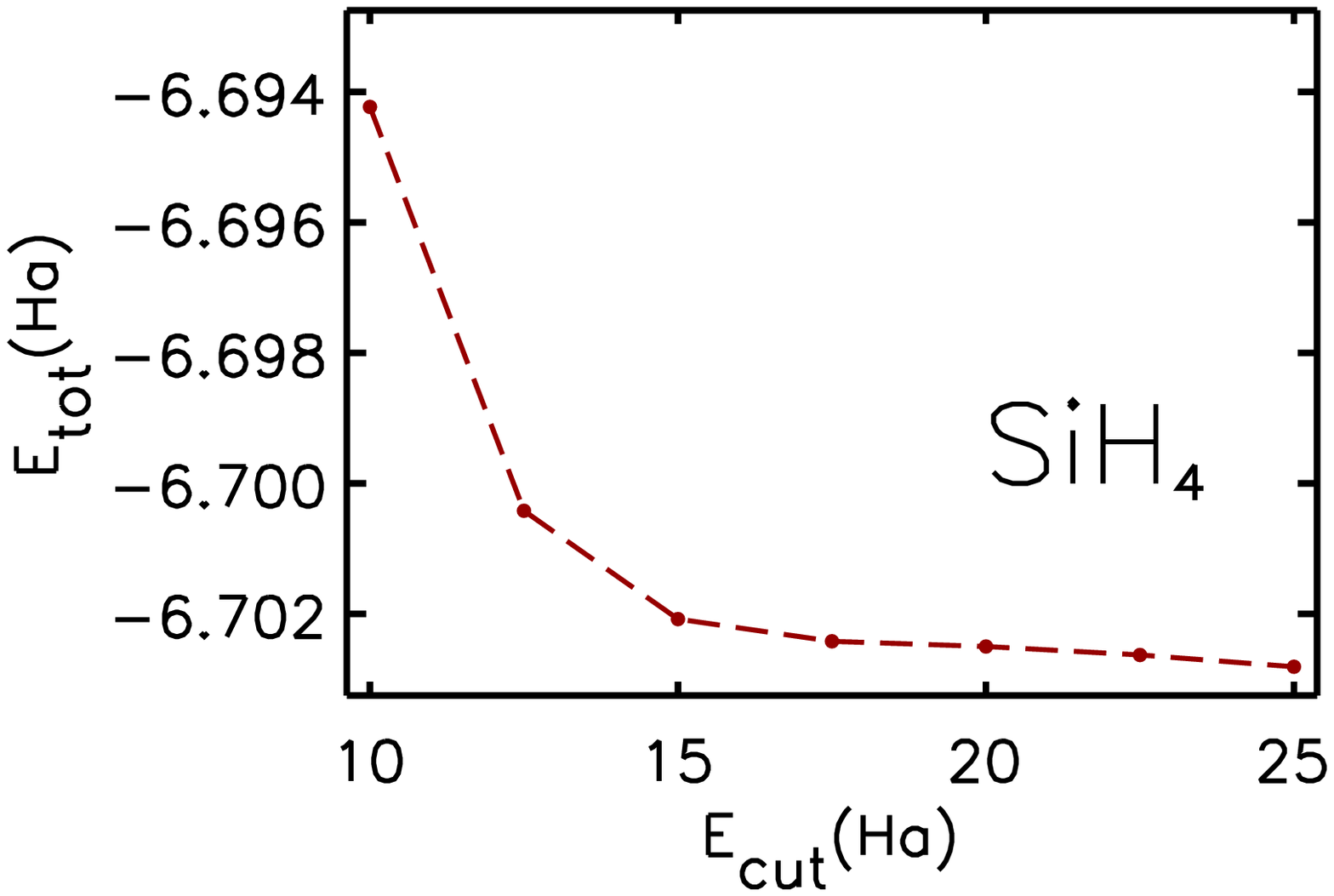}
\end{minipage}%\hspace{1pc}%
\begin{minipage}{12.5pc}
\includegraphics[width=12.5pc]{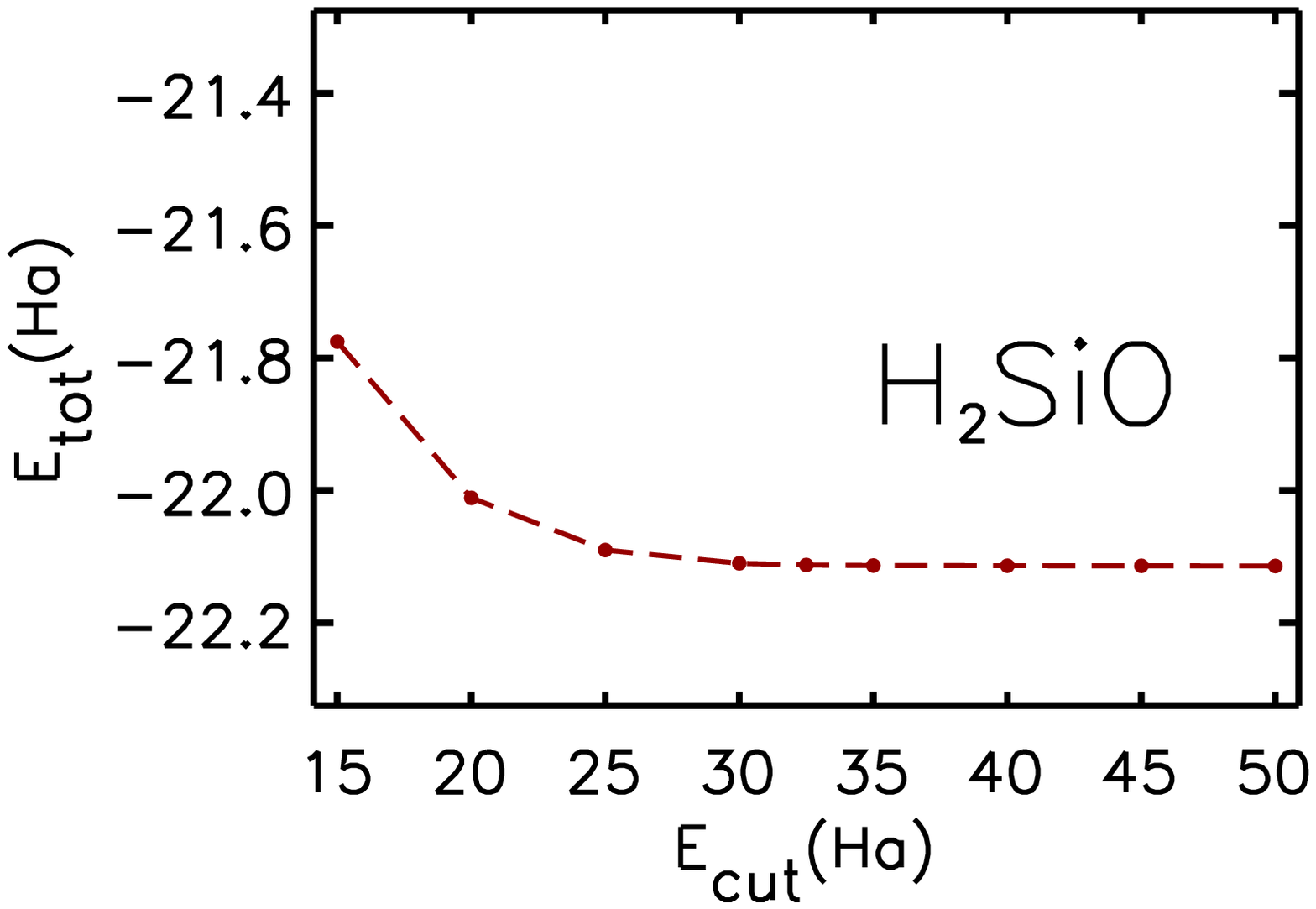}
\end{minipage} 
\caption{\label{ConvSmall}
  Total energy $E_{\rm tot}$as a function of the supercell lattice
  parameter $a_{\rm cell}$ for silan (left), where the total energy is
  reduced by an offset of 6702 mHa, as a function of the kinetic
  energy cutoff $E_{\rm cut}$ for silan (middle) and silanon
  (right). Note that the scale of the total energy in the left figure
  is in mHa, whereas the scale in the other two figures is in Ha.  }
\end{figure}

The results for the variation of the total energy $E_{\rm tot}$ as a
function of the of the supercell lattice parameter ($a_{\rm cell}$)
and as a function of the $E_{\rm cut}$ are shown in
Fig.~\ref{ConvSmall}. We have performed the convergence tests with
respect to the $E_{\rm cut}$ for both silan and silanon, since the
description of the oxygen requires a larger number of plane waves than
for the other atoms. As visible in the figure, the size of the
supercell has a rather small influence to the calculation. The
variation of $E_{\rm tot}$ is in the $\mu$Ha range and therefore we
have chosen $a_{\rm cell}=30~{\rm a_B}$. The variation of $E_{\rm
tot}$ with respect to the number of plane waves is larger. For silan
convergence has been achieved at $E_{\rm cut}=17.5~{\rm Ha}$ yielding
an error of less than 0.4~mHa in $E_{\rm tot}$. Compared with silan
the total energy of silanon converges a factor of 10 slower. At
$E_{\rm cut}=32.5~{\rm Ha}$ we found a difference of less than 1.5~mHa
compared with the value of $E_{\rm tot}$ at absolute
convergence. Thus, $E_{\rm cut}=32.5~{\rm Ha}$ is sufficient for the
numerical description of silanon.

Since we want to investigate the vibrational excitations we inspected
also the frequencies of silan and silanon. The lowest six frequencies,
the translational and rotational ones should vanish. Due to the
numerical noise and incompleteness of the plane-wave basis set, these
frequencies do not vanish exactly. However, in the case of silanon,
which is less converged with respect to the number of plane waves,
they have values less than 18${\rm cm}^{-1}$ which is sufficiently
small. For silan they are even smaller.
%%%%%%%%
%
\subsection{Eigenvectors for silan and silanon} \label{SectSmallEig}
\begin{figure}[b]
  \begin{minipage}{7.7pc}
    \includegraphics[height=7pc,angle=0]{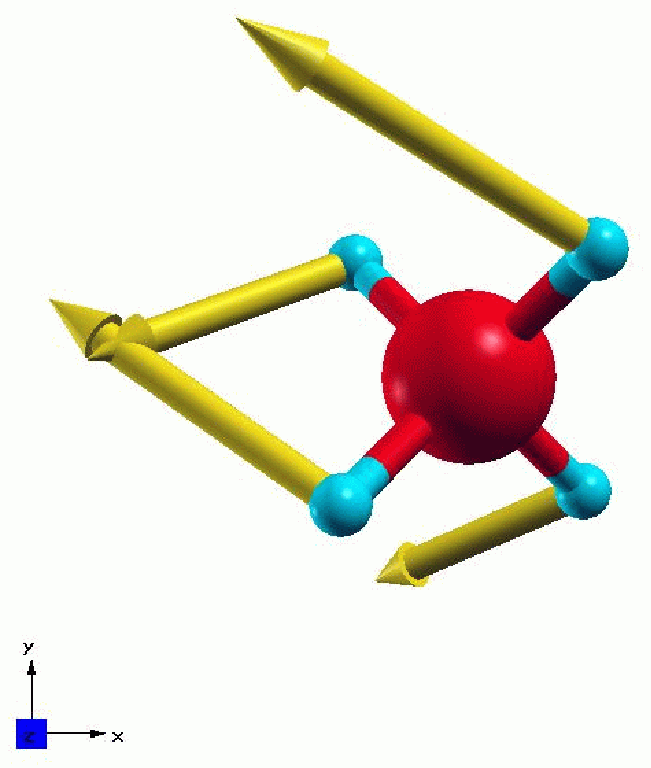}\\
    859~cm$-1$ \\ scissor asymmetric
  \end{minipage}\hspace{2pc}
  \begin{minipage}{7.7pc}
    \includegraphics[height=7pc,angle=0]{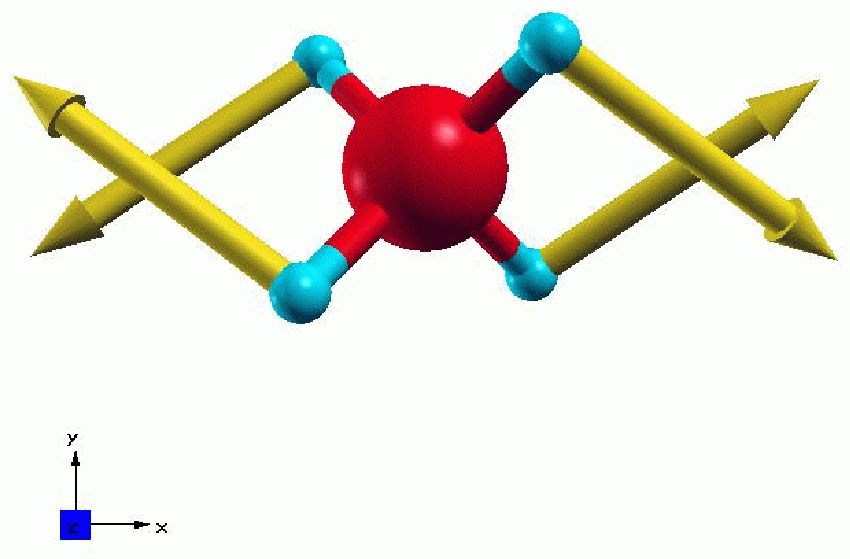}\\
    941~cm$-1$ \\ scissor symmetric
  \end{minipage}\hspace{2pc}
  \begin{minipage}{7.7pc}
    \includegraphics[height=7pc,angle=0]{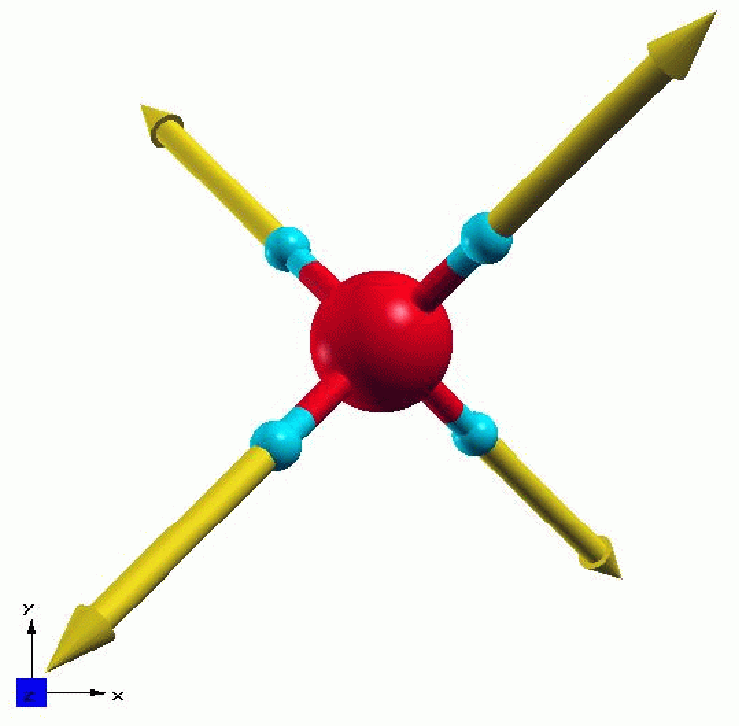}\\
    2147~cm$-1$ \\ stretching sym.
  \end{minipage}\hspace{2pc}
  \begin{minipage}{7.7pc}
    \includegraphics[height=7pc,angle=0]{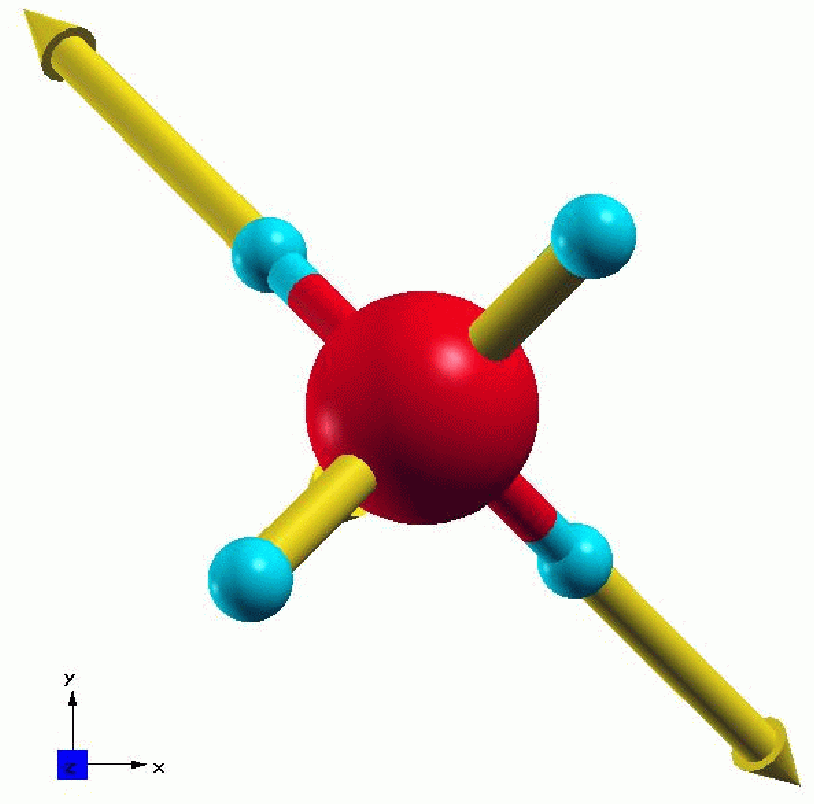}\\
    2171~cm$-1$ \\ stretching asym.
  \end{minipage}
  \caption{\label{EigenSilan}
    Eigenvectors of silan with symmetric (sym) and asymmetric (asym)
    displacements for the frequencies denoted in the subsets. For the
    assignment of the atoms see Fig.~\ref{StructSmall}. }
\end{figure}

The vibrations of silan and silanon can be characterized by their
displacement patterns. Silan has 15 vibrational modes where 6 are
vanishing. Analyzing the frequencies, we found that the lowest
frequency is threefold, the next twofold, and the highest threefold
degenerated due to the symmetry of the system. The eigendisplacements
are shown in Fig.~\ref{EigenSilan}. There are two scissor and two
stretching patterns, where in each case there is a symmetric
displacement and an asymmetric one.

\begin{figure}[t]
  \begin{minipage}{8pc}
    \includegraphics[height=7pc,angle=0]{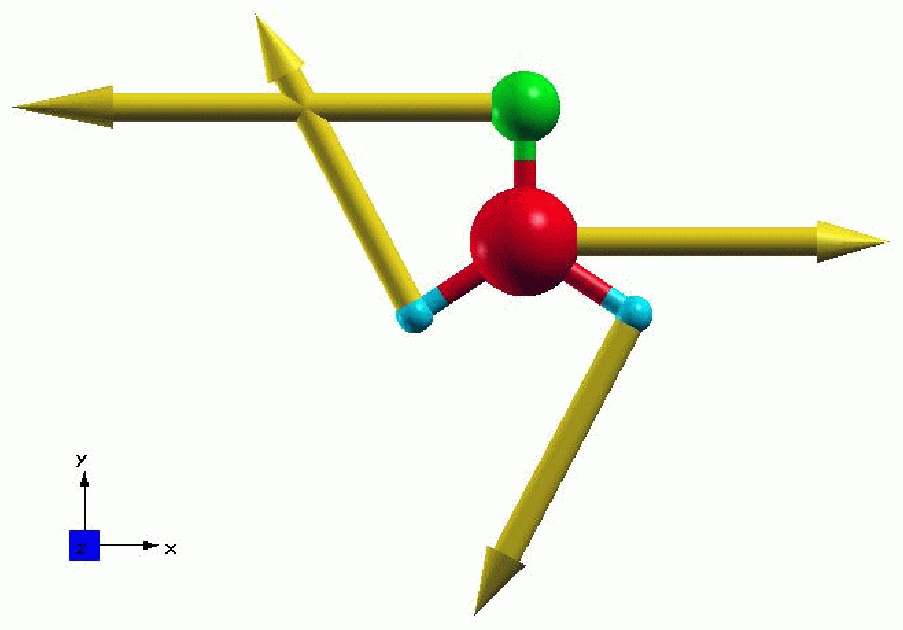}\\
    652~cm$-1$ \\ bending x-y
  \end{minipage}\hspace{5pc}
  \begin{minipage}{8pc}
    \includegraphics[height=7pc,angle=0]{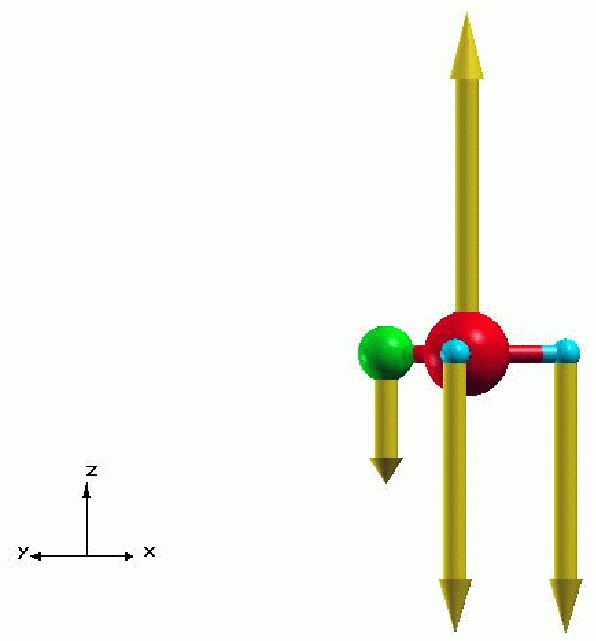}\\
    657~cm$-1$ \\ bending z
  \end{minipage}\hspace{5pc}
  \begin{minipage}{8pc}
    \includegraphics[height=7pc,angle=0]{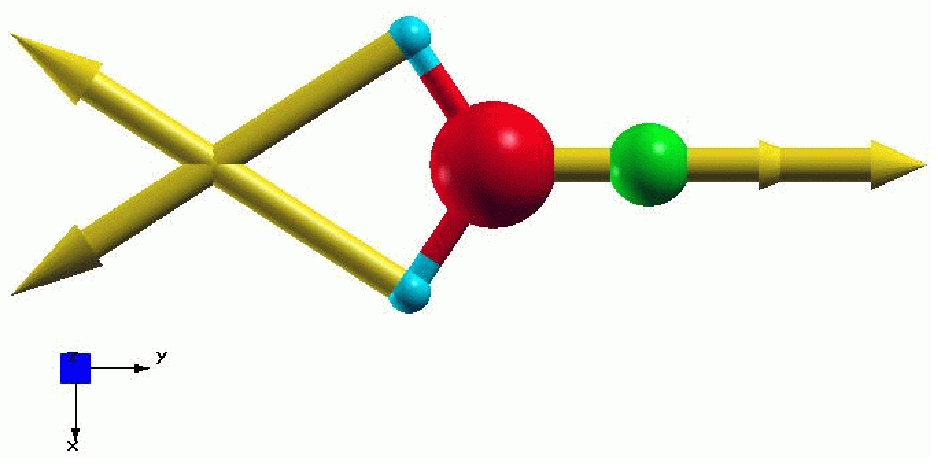}\\
    962~cm$-1$ \\ scissor bending
  \end{minipage}

  \begin{minipage}{8pc}
    \vspace{1pc}
    \includegraphics[height=5pc,angle=0]{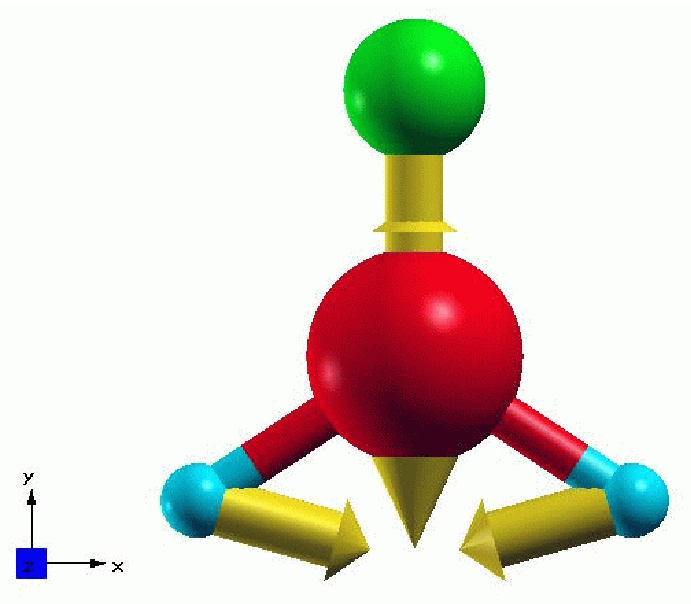} \vspace{1pc}\\
    1196~cm$-1$ \\ stretching Si=O
  \end{minipage}\hspace{5pc}
  \begin{minipage}{8pc}
    \includegraphics[height=7pc,angle=0]{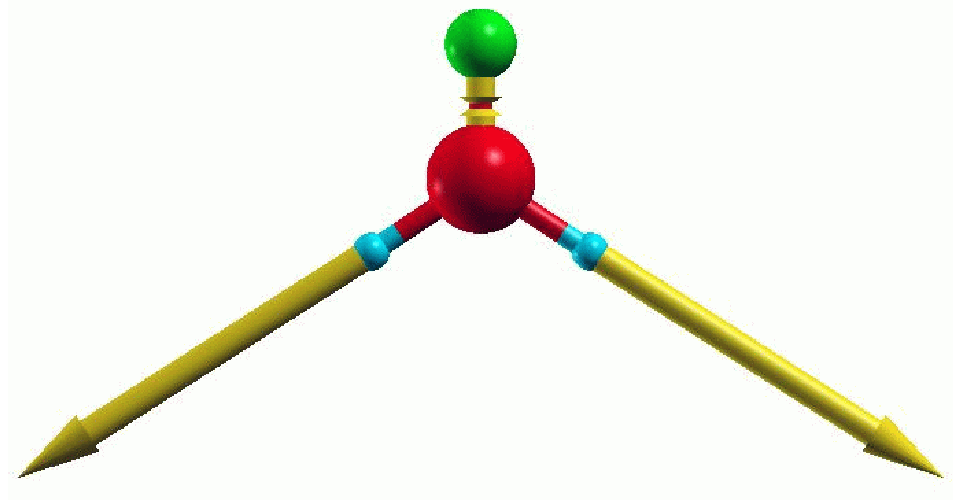}\\
    2111~cm$-1$ \\ stretch. sym. Si-H
  \end{minipage}\hspace{5pc}
  \begin{minipage}{8pc}
    \includegraphics[height=7pc,angle=0]{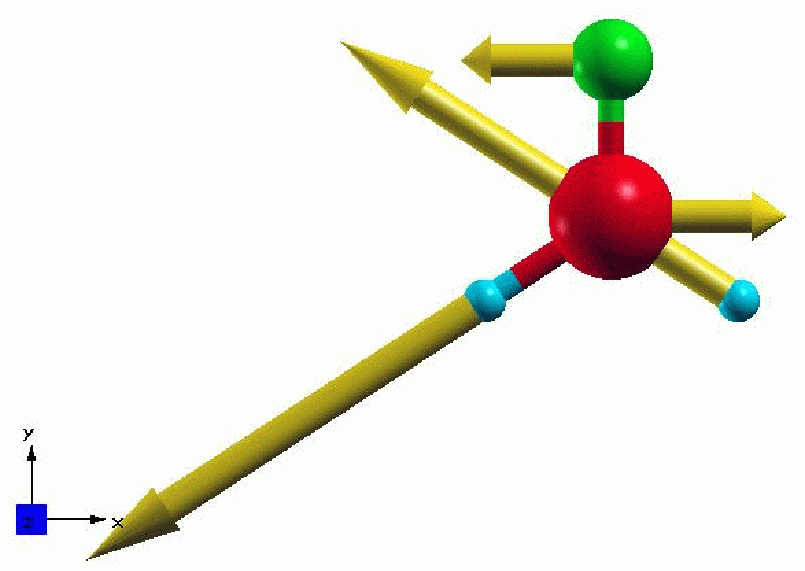}\\
    2147~cm$-1$ \\ stretch. asym. Si-H
  \end{minipage}

  \caption{\label{EigenSilanon}
    Eigenvectors of silanon with symmetric (sym) and asymmetric (asym)
    displacements for the frequencies denoted in the figures. The
    length of the oxygen- and silicon-related eigenvectors have been
    expanded for visibility. For the assignment of the atoms see
    Fig.~\ref{StructSmall}
  }
\end{figure}

For silanon we have in total 12 frequency eigenvalues, from these six
are vanishing. The remaining six true vibrations show no
degeneracy. As for silan we have inspected the eigenvectors of silanon
which are displayed in Fig.~\ref{EigenSilanon}. Besides the two
scissor and the two stretching modes, there are two bending modes,
where for one the displacement of the hydrogen atoms is in the x-y
plane, and for the other it is in the z direction (Note: the x-y plane
is defined as the plane of the planar molecule). Also here, we have
found symmetric and asymmetric modes for the scissor and the
stretching mode. The stretching modes can be divided in two classes:
modes with a large movement of the oxygen and modes where the oxygen
is displaced only a little. Usually the displacements of the hydrogens
are at least a factor of 10 larger than the ones of the other
atoms. Therefore, we have rescaled the eigenvectors of oxygen and
silicon in the figure for visibility. There is just one mode where the
displacement of the oxygen is in the same order of magnitude as the
hydrogens: the stretching Si=O mode. This vibration mode is
characteristic for the presence of oxygen in silanon.

%%%%%%%%
%
\subsection{Comparison with experimental results}\label{SectSmallExp}

\begin{table}[b]
\lineup
\caption{\label{CompExpSilan} Calculated bond distance and vibrational
  frequencies for silan in comparison with experimental results from
  of (a) Cadorna \cite{Car1983} and (b) Boyd \cite{Boy1955} and
  theoretical tight-binding data.}
\begin{center}
\begin{tabular}{lllll}
\br
Silan & this work &  th \cite{Kim1994} & th \cite{Min1992} &  Experiment \\
\mr
Distance Si-H & 1.4872 \AA & 1.48 \AA & 1.48 \AA &  1.4798$^b$ \AA \\
\mr
\multicolumn{5}{l}{Frequencies (degeneracy):} \\
\mr
Scissor asym. (3)         &\0859 cm$^{-1}$&\0871 cm$^{-1}$&\0831 cm$^{-1}$ &\0911$^a$ cm$^{-1}$ \\
Scissor sym. H-Si-H (2)   &\0941 cm$^{-1}$&\0976 cm$^{-1}$&\0984 cm$^{-1}$ &\0976$^a$ cm$^{-1}$ \\
Stretching sym. H-Si (1)  & 2147 cm$^{-1}$& 2226 cm$^{-1}$& 2226 cm$^{-1}$ & 2178$^a$ cm$^{-1}$ \\
Stretching asym. H-Si (3) & 2171 cm$^{-1}$& 2291 cm$^{-1}$&                & 2191$^a$ cm$^{-1}$ \\
\br
\end{tabular}
\end{center}
\end{table}
For the small clusters silan and silanon there are experimental and
theoretical reference data available to which we can compare our
results. With this comparison we can also prove the reliability of the
computational approach used here.

A comparison of our results for silan with the measured data and
available theoretical data is presented in
Tab.~\ref{CompExpSilan}. The overall agreement is good, however the
experimental frequencies are slightly underestimated by the calculated
ones. The relative difference between the results for the frequencies
of Cardona \cite{Car1983} and ours is 6~\% for the asymmetric scissor
mode, 4~\% for the symmetric one, 1~\% for the symmetric and the
asymmetric stretching mode. Besides, the agreement of the
silicon-hydrogen bond length is excellent. For the lower frequencies
the tight-binding results are more close to the experimental
frequencies, however, these calculations used experimental values for
the fitting of the tight-binding parameters whereas our calculation is
fully ab initio.

For silanon there are just two experimental values available which
have been obtained by infrared spectroscopy of silanon in an Ar matrix
\cite{Wit1985}: the silicon-oxygen stretching frequency at
1202~cm$^{-1}$ and a frequency at 697~cm$^{-1}$ where the assignment
is not clear. It corresponds either to the bending frequency in x-y
plane or to the one in the z direction.  There exist theoretical
investigations of the vibrational properties of silanon, based
ab-initio approaches like molecular-orbital theory, Hartree-Fock,
Configuration Interaction, Coupled-Cluster methods, and others
\cite{Har2004,Gol1998b,Dar1993,Kud1984,Mar1998,Ma1994,Kop1999}.  In
these calculations the results vary depending on the method, even
sometimes within the same theoretical approach, e.g., just changing
the basis set used in the local-orbital expansion. For example, the
Si=O stretching frequency has been obtained at 1355~cm$^{-1}$ by
Darling and Schlegel \cite{Dar1993}, 1182~cm$^{-1}$ by Gole and Dixon
\cite{Gol1998b}, and 1217~cm$^{-1}$ by Hargittai and R{\'e}ffy
\cite{Har2004} all by using the same molecular-orbital theory
implementation in GAUSSIAN \cite{GAUSSIAN}. Thus, we have compared our
results with some of the most recent theoretical values, which are the
the molecular-orbital theory results using a B3LYP/6-311G(d,p) DFT
basis set of Hargittai and R{\'e}ffy \cite{Har2004}, the
molecular-orbital theory results using triple $\zeta$ valence basis
set at the local DFT level of Gole and Dixon \cite{Gol1998b}, and the
coupled cluster method results using quadruple $\zeta$ basis set of
Martin \cite{Mar1998}. As visible in Tab.~\ref{CompExpSilanon}, the
agreement with the theoretical references is very good. Therefore, the
vibrational properties of isolated systems can be described very well
using a periodic-cell approach based on plane waves instead on local
orbitals. Inspecting the results more detailed, our results are very
close to the ones of Gole and Dixon \cite{Gol1998b} also using a
density-functional theory approach. Our results are in parts more
close to the experimental values than the frequencies calculated by
other groups.

\begin{table}[b]
\lineup
\caption{\label{CompExpSilanon} Calculated bond distances and
  vibrational frequencies for silanon in comparison with experimental
  (exp) results from Ref~\cite{Bog1996} (a) and Ref~\cite{Wit1985} (b)
  and theoretical (th) ones.}
\begin{center}
\begin{tabular}{llllll}
\br
Silanon&this work&th. \cite{Har2004}&th. \cite{Gol1998b}&th. \cite{Mar1998}& exp.\\
\mr
\multicolumn{5}{l}{Distances (\AA)} \\
\mr
Si-H & 1.491 & 1.482 & 1.505  & 1.478 & 1.472$^a$\\
Si=O & 1.505 & 1.517 & 1.534  & 1.522 & 1.515$^a$\\
\mr
\multicolumn{5}{l}{Frequencies (cm$^{-1}$)} \\
\mr
Bending x-y plane      &\0652 &\0706 &\0676 & 680 &\0697$^b$ \\
Bending z direction    &\0657 &\0712 &\0693 & 692 & \\
Scissor bending H-Si-H &\0962 & 1023 &\0977 & 993 & \\
Stretching Si=O        & 1196 & 1217 & 1197 &1203 & 1202$^b$ \\
Stretching sym. H-Si   & 2111 & 2223 & 2127 &2162 &  \\
Stretching asym.H-Si   & 2140 & 2238 & 2132 &2186 &  \\
\br
\end{tabular}
\end{center}
\end{table}

Last, we can compare the vibrations of silan with those of
silanon. Here we found that the frequency of the Si-H-Si
scissor-bending, the symmetric and the asymmetric Si-H modes in both
systems have very similar frequencies which differ less than 2~\%
between silan and silanon. Thus, the intermediate Si=O stretching
frequency can be utilized to characterize silanon experimentally.

%
%%%%%%%%%%%%%%%%%%%%%%%%%%%%%%%%%%%%%%%%%%%%%%%%%%%%%%%%%%%%%%%%%%%%%
%
\section{Si$_{10}$H$_{16}$ and Si$_{10}$H$_{14}$O}\label{SectBig}

\begin{figure}[t]
\begin{minipage}{16pc}
\includegraphics[width=7.7pc]{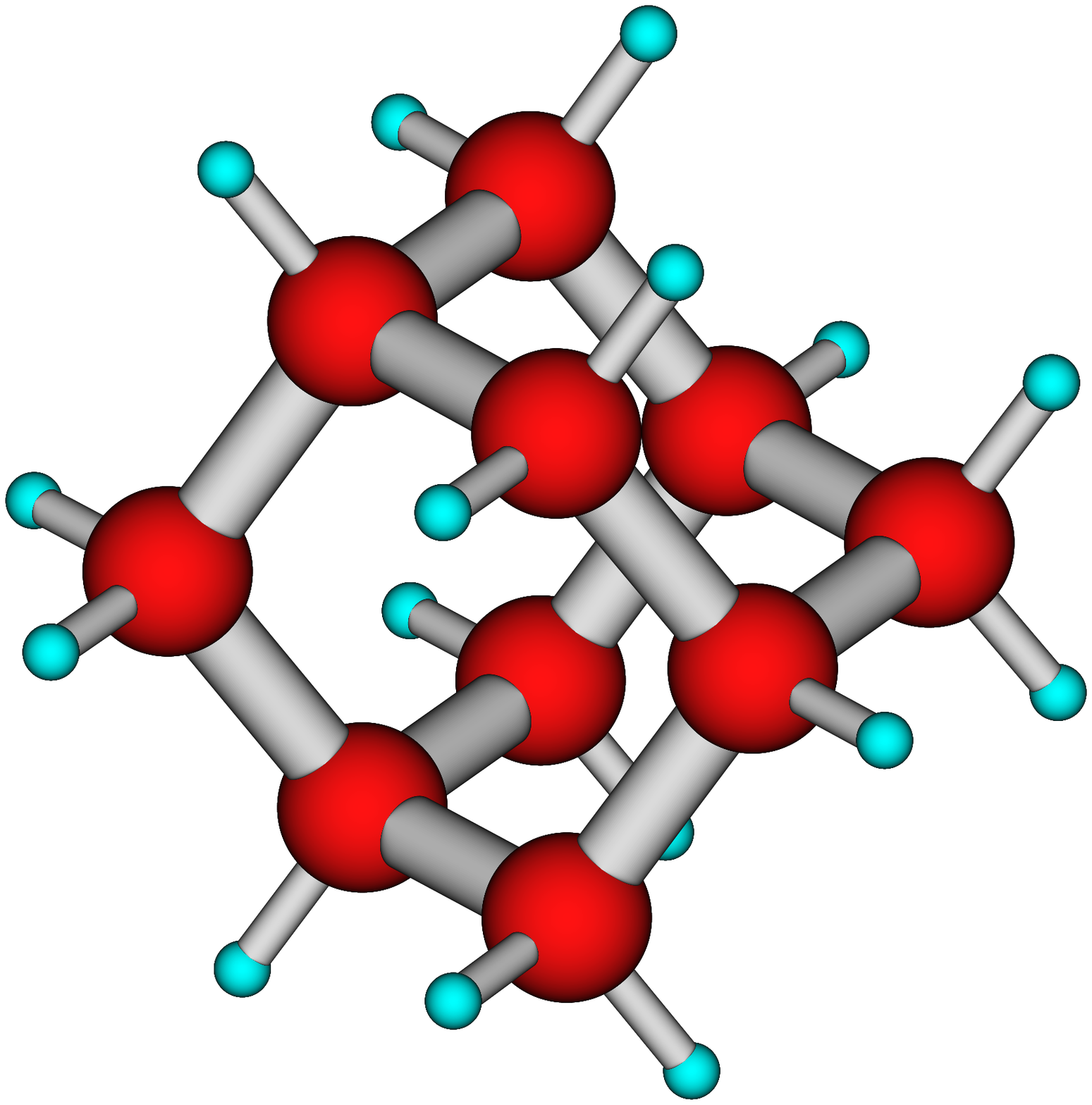}
\includegraphics[width=7.7pc]{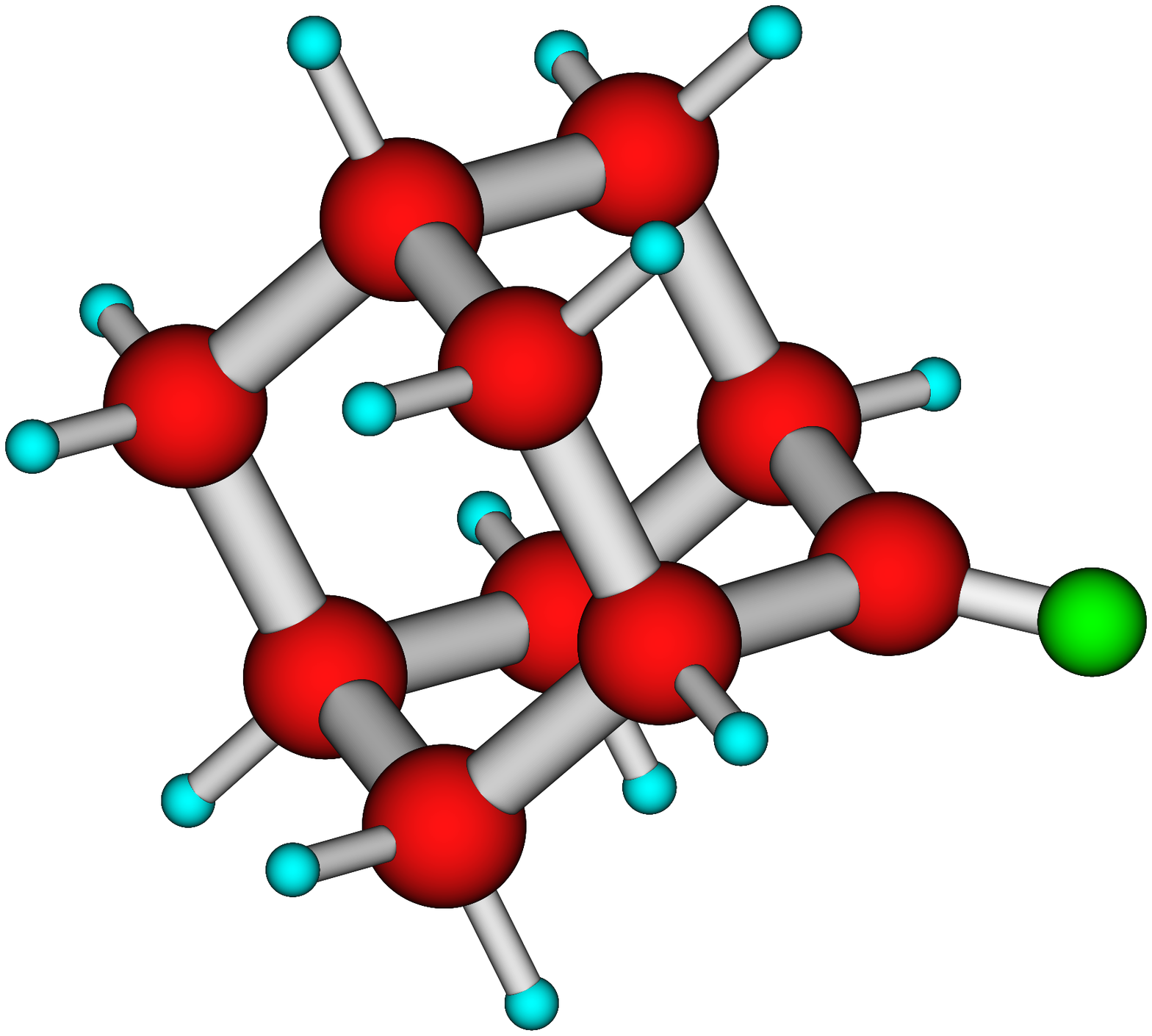}
\end{minipage}\hspace{2pc}%
\begin{minipage}{16pc}
\caption{\label{StructBig}
  Si$_{10}$H$_{16}$ (left) and Si$_{10}$H$_{14}$O (right). The
  assignment of the atoms is as in Fig.~\ref{StructSmall}.}
\end{minipage} 
\end{figure}

The big nanocrystals Si$_{10}$H$_{16}$ and Si$_{10}$H$_{14}$O have
been investigated in a similar way as the small ones.
Si$_{10}$H$_{16}$ has the same space group symmetry as silan. It is
displayed in Fig.~\ref{StructBig}. In Si$_{10}$H$_{16}$ there are two
kinds of silicon atoms: those which are bonded to three other silicon
atoms and which have therefore only one dangling bond which is
saturated with a hydrogen atom, and those which are bonded to two
other silicon atoms where two hydrogen atoms are necessary to saturate
the bonds. There are various possibilities to oxidize this
crystal. The oxygen can be bonded between first or second-neighbored
silicon atoms, where each of the oxygen-bonding silicons will loose
one hydrogen. However, we have chosen another configuration for
Si$_{10}$H$_{14}$O (see Fig.~\ref{StructBig}) where the oxygen is
double-bonded to a silicon atom yielding the same space group symmetry
as silanon. Even if this configuration is not the energetically most
favorable one, it is just $\approx$60~mHa less stable than the most
stable one. Since H$_2$SiO and Si$_{10}$H$_{14}$O have both a
double-bonded oxygen, one can compare the corresponding Si=O
stretching frequencies.

Before investigating the vibrational properties of Si$_{10}$H$_{16}$
and Si$_{10}$H$_{14}$O we want to reinspect the convergence parameters
(Sect.~\ref{SectConvBig}). Afterwards, we analyze the vibrational
density of states (Sect.~\ref{SectDos}) and the displacement patterns
for oxygen-related modes (Sect.~\ref{SectBigEigen}).
%
%%%%%%%%
%
\subsection{Convergence}\label{SectConvBig}

\begin{table}[b]
\lineup
\caption{\label{ConvSilanon} 
  Variation of the vibrational frequencies in ${\rm cm^{-1}}$ of
  silanon for various values of the kinetic energy cutoff $E_{\rm
  cut}$.}
\begin{center}
\begin{tabular}{lllllll}
\br
  ${\rm E}_{\rm cut}$~(Ha)&15 & 20 & 25 & 30 & 32.5 & 35 \\
\mr
  Bending (x-y plane)    &\0638 &\0647 &\0653 &\0651 &\0652 &\0652 \\
  Bending (z dir.)       &\0648 &\0655 &\0657 &\0657 &\0657 &\0657 \\
  Scissor bending H-Si-H &\0960 &\0961 &\0962 &\0961 &\0962 &\0962 \\
  Stretching Si=O        & 1150 & 1183 & 1195 & 1196 & 1196 & 1196 \\
  Stretching sym. H-Si   & 2109 & 2107 & 2110 & 2110 & 2111 & 2111 \\
  Stretching asym.H-Si   & 2137 & 2137 & 2140 & 2140 & 2140 & 2141 \\
\br
\end{tabular}
\end{center}
\end{table}

Since Si$_{10}$H$_{16}$ and Si$_{10}$H$_{14}$O are larger than SiH$_4$
and H$_2$SiO it is necessary to increase the size of the supercell
from $a_{\rm cell}=30~{\rm a_B}$ for the small clusters to $a_{\rm
cell}=40~{\rm a_B}$ for the big ones. With this choice, the
kinetic-energy cutoff of $E_{\rm cut}=17.5~{\rm Ha}$ can easily be
used for the non-oxidized nanocrystal, whereas $E_{\rm cut}=32.5~{\rm
Ha}$ for the oxidized one is beyond the computational limits. Thus, we
look at the vibrational frequencies for silanon as a function of
$E_{\rm cut}$ (see Tab.~\ref{ConvSilanon}). As noticed from the table,
most of the frequencies do not vary significantly with the number of
plane waves already from medium values of $E_{\rm cut}$. The largest
variations are found for the stretching Si=O frequency. However, even
this frequency is already converged at $E_{\rm cut}=25~{\rm
Ha}$. Thus, we could reduce the kinetic energy cutoff for silanon
without obtaining significantly different results. In order to see if
this conclusion is also true for Si$_{10}$H$_{14}$O, we have
calculated the the vibrational spectra at $E_{\rm cut}=15~{\rm Ha}$
and at $E_{\rm cut}=25~{\rm Ha}$. For the Stretching Si=O we have
found a difference of 52~${\rm cm^{-1}}$ between these two
calculations, which is nearly the same as for silanon comparing the
values at the same kinetic-energy cutoffs. Since the other frequencies
of Si$_{10}$H$_{14}$O did not vary significantly between the
calculations using $E_{\rm cut}=15~{\rm Ha}$ and $E_{\rm cut}=25~{\rm
Ha}$, we can assume that convergence has been achieved at 25~Ha. This
finding was confirmed, since the vanishing frequencies are lower than
16~${\rm cm}^{-1}$.

%
%%%%%%%%
%
\subsection{Vibrational density of states} \label{SectDos}
\begin{figure}[t]
\includegraphics[width=18pc]{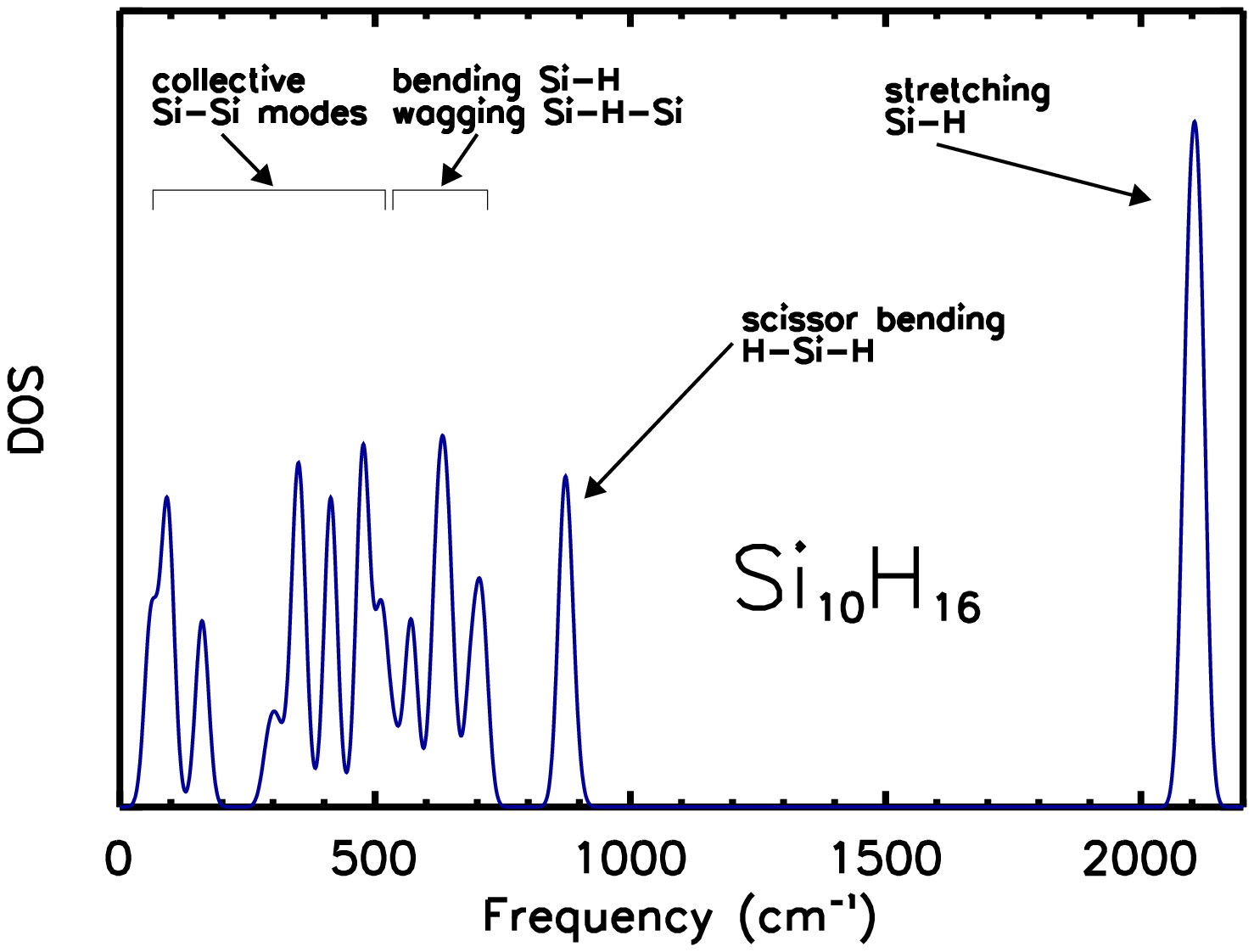}\hspace{1.8pc}%
\includegraphics[width=18pc]{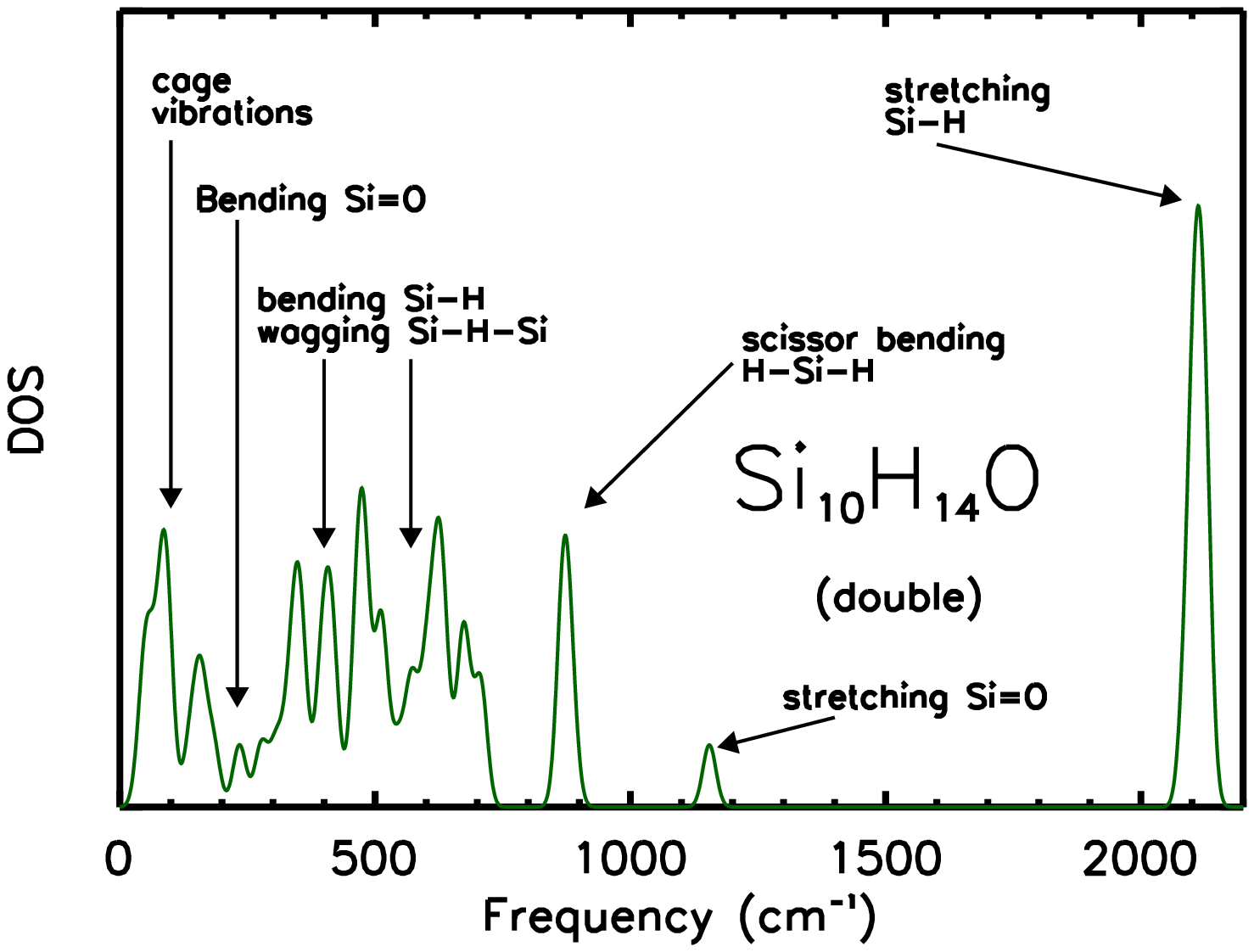}%
\caption{\label{DOS}
  Vibrational density of states (DOS) of Si$_{10}$H$_{16}$ (left) and
  Si$_{10}$H$_{14}$O (right) together with the assignment of the modes.}
\end{figure}

For the big nanocrystals we obtained a large number of frequencies. In
the case of Si$_{10}$H$_{16}$ we have found 78 frequency eigenvalues,
where 6 frequencies vanish. However, most of the modes are twofold or
threefold degenerated. For Si$_{10}$H$_{14}$O there are 75 frequencies
and non of them is degenerated. Thus, we have calculated the
vibrational density of states (DOS), where we applied a Gaussian
broadening of a width of 12.8~${\rm cm^{-1}}$. The result is presented
in Fig.~\ref{DOS}. To our knowledge, there are no experimental data
available for comparison for these two clusters.

Analyzing the DOS for Si$_{10}$H$_{16}$, we can distinguish four
regions: the low-frequency collective Si-Si modes with frequencies up
to $\approx 500~{\rm cm^{-1}}$, from $\approx 500~{\rm cm^{-1}}$ to
$\approx 750~{\rm cm^{-1}}$ the Si-H-Si bending and wagging modes, the
peak at $\approx 870~{\rm cm^{-1}}$ corresponds to the Si-H-Si
scissor-bending modes, and at $\approx 2100~{\rm cm^{-1}}$ we have the
Si-H stretching modes. Compared to silan, the Si-H stretching modes
have slightly lower frequencies.

In the DOS of Si$_{10}$H$_{14}$O we find the same regions for the Si
and Si-H related vibrations. In addition, we find in the frequency
gaps some oxygen-related vibrations: at $234~{\rm cm^{-1}}$ a Si=O
bending mode and at $1155~{\rm cm^{-1}}$ the Si=O stretching
mode. These peaks in the DOS are the oxygen-signature in the
spectra. The Si=O stretching frequency is comparable to that of
silanon, which was at $1196~{\rm cm^{-1}}$. For the Si=O bending mode,
the displacement of Si and O is similar to the bending x-y mode of
silanon, however, it is at a quite different frequency.

We have compared also the DOS of Si$_{10}$H$_{16}$ with the one of
Si$_{10}$H$_{14}$O. Here we observe that the peak of the Si-H
stretching modes and the one of the Si-H-Si scissor-bending modes
coincidence for both clusters and thus, they are independent of the
presence of the oxygen. In contrary, the regions of the Si and Si-H
related frequencies of Si$_{10}$H$_{16}$ are slightly different from
the ones of Si$_{10}$H$_{14}$O, since in the latter case there is an
overlap with oxygen related modes. Nevertheless, it would be possible
to discriminate experimentally the oxidized nanocrystals from the
non-oxidized one due to the characteristic Si=O frequencies.

%%%%%%%
%
\subsection{Analysis of displacement patterns}\label{SectBigEigen}
\begin{figure}[t]
\begin{minipage}{24pc}
\includegraphics[width=10pc]{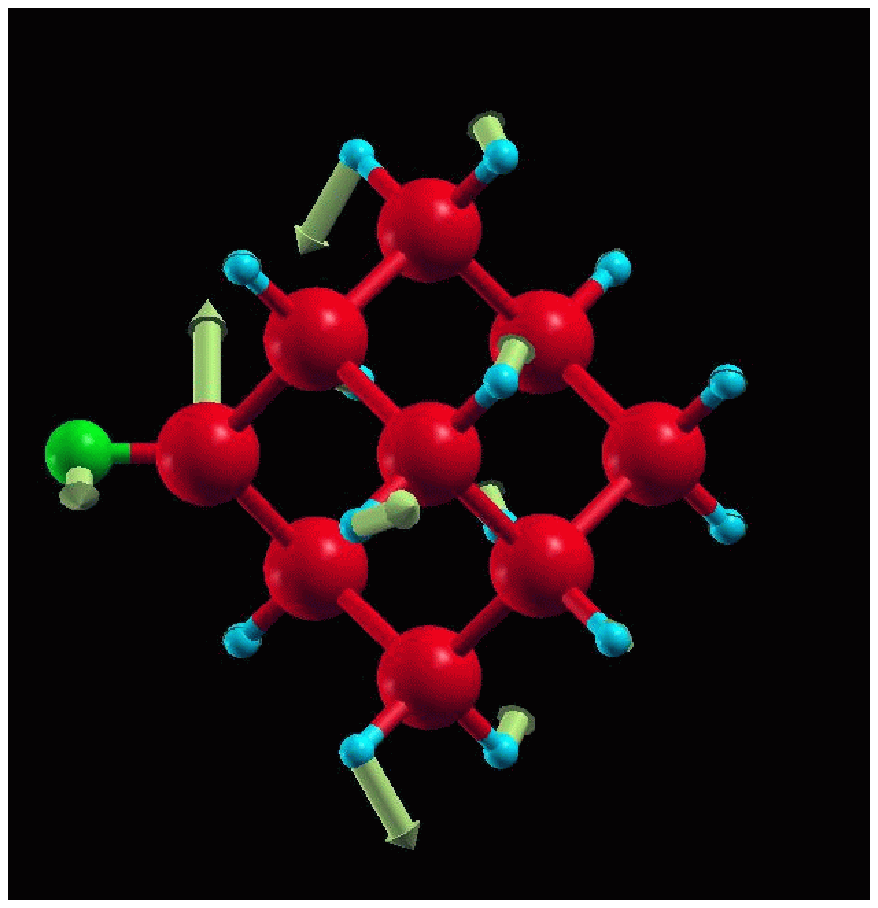}\hspace{2pc}%
\includegraphics[width=10pc]{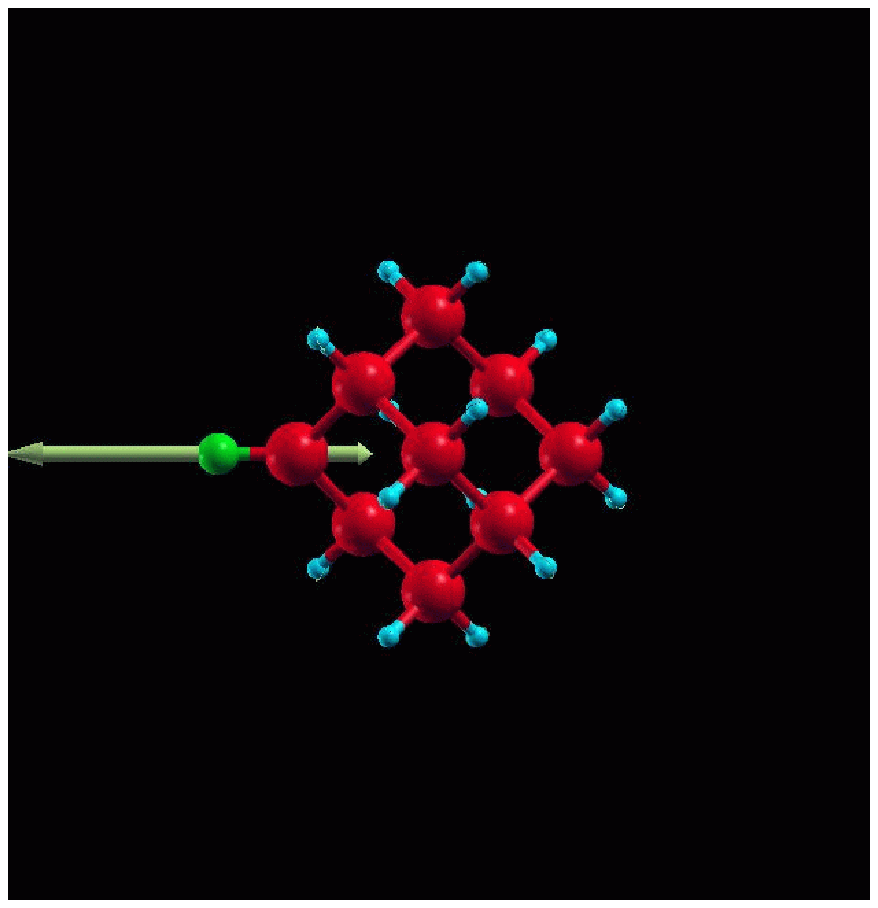}\hspace{2pc}%
\end{minipage}
\begin{minipage}{13pc}
\caption{\label{EigenBig} 
  Eigendisplacements of oxygen-related modes, Si=O bending (left) and
  Si=O stretching (right) modes the frequencies 234~${\rm cm^{-1}}$
  and 1155~${\rm cm^{-1}}$, respectively.}
\end{minipage}
\end{figure}

After we have identified the characteristic oxygen-related spectral
features in the DOS, we also want to analyze the displacement patterns
of the corresponding modes. The eigenvectors of the Si=O bending and
the Si=O stretching modes are displayed in Fig.~\ref{EigenBig}. The
Si=O bending mode has a delocalized character, where the Si=O
stretching one is completely localized at the oxygen and its
neighboring silicon atom. Both modes show no inversion symmetry and
are Raman active.

%
%%%%%%%%%%%%%%%%%%%%%%%%%%%%%%%%%%%%%%%%%%%%%%%%%%%%%%%%%%%%%%%%%%%%%
%
\section{Summary and outlook}
We have investigated the silicon nanocrystals SiH$_4$ (silan),
H$_2$SiO (silanon), Si$_{10}$H$_{16}$, and Si$_{10}$H$_{14}$O with
respect to their vibrational properties. The vibrational frequencies
of silan and silanon are in good agreement with experimental and
theoretical reference data. This shows, that localized systems like
nanocrystals are well described using a plane-wave approach within
periodic-boundary conditions.  We have computed the vibrational
density of states for the non-oxidized Si$_{10}$H$_{16}$ and the
corresponding oxidized one Si$_{10}$H$_{14}$O where the oxygen is
double-bonded to a silicon atom. The frequencies have been analyzed
with respect to their vibrational character. For Si$_{10}$H$_{16}$ we
have found four regions related to Si-Si vibrations and various Si-H
motions. The same regions are present also in the density of states of
Si$_{10}$H$_{14}$O cluster, where we have found additional peaks in
the frequency gaps of Si$_{10}$H$_{16}$. These additional peaks have
been identified as a non-localized Si=O bending mode and a localized
Si=O stretching mode.  Thus, comparing the vibrational density of
states of the oxidized and the non-oxidized nanocrystal, we find a
clear signature of the oxygen in the spectra.

A further investigation of these nanocrystals will contain an analysis
of the vibrational spectra of other Si$_{10}$H$_{14}$O isomers, as well
as a characterization of localized and non-localized modes and their
Raman activity.

%
%%%%%%%%%%%%%%%%%%%%%%%%%%%%%%%%%%%%%%%%%%%%%%%%%%%%%%%%%%%%%%%%%%%%%
%
\ack
This work was funded in part by the EU's 6th Framework Programme
through the NANOQUANTA Network of Excellence
(NMP-4-CT-2004-500198). Computer facilities at CINECA granted by INFM
(Project no. 643/2006) are gratefully acknowledged. We also like to
thank Matteo Gatti, Stefano Ossicini, and Paolo Giannozzi for fruitful
discussions.

%%%%%%%%%%%%%%%%%%%%%%%%%%%%%%%%%%%%%%%%%%%%%%%%%%%%%%%%%%%%%%%%%%%%%
%%%%%%%%%%%%%%%%%%     End of the paper                %%%%%%%%%%%%%%
%%%%%%%%%%%%%%%%%%%%%%%%%%%%%%%%%%%%%%%%%%%%%%%%%%%%%%%%%%%%%%%%%%%%%
\section*{References}
\bibliography{Theory,SiNC}

\end{document}